\documentclass[aps, prbr, amsmath,amssymb, reprint, showpacs,superscriptaddress,groupedaddress]{revtex4-1}

\usepackage{graphicx}
\usepackage{dcolumn}
\usepackage{bm}
\usepackage{amsmath}
\usepackage{color}

\usepackage{dcolumn}   
\usepackage{amsmath}
\usepackage{multirow}
\usepackage{graphicx}    
\usepackage{subfigure}
\usepackage{comment}
\usepackage{color}
\usepackage[colorlinks,bookmarks=false,citecolor=blue,linkcolor=red,urlcolor=blue]{hyperref}
\usepackage{nicefrac}
\usepackage{soul}

\usepackage{nicefrac}

\begin{document}

	\title{Anomalous random correlations of force constants on the lattice dynamical properties of disordered Au-Fe alloys}
	\author{Jiban Kangsabanik,$^1$ Rajiv K. Chouhan,$^2$ D.~D. Johnson,$^{2,3}$ Aftab Alam$^1$}
	\affiliation{$^1$Department of Physics, Indian Institute of Technology Bombay, Mumbai 400076, India}
	\affiliation{$^2$The Ames Laboratory, U.S. Department of Energy, Ames, Iowa 50011, USA} 
	\affiliation{$^3$Materials Science \& Engineering, Iowa State University, Ames, Iowa 50011, USA}

	\begin{abstract}
		Au-Fe alloys are of immense interest due to their biocompatibility, anomalous hall conductivity, and applications in various medical treatment. However, irrespective of the method of preparation, they often exhibit a high-level of disorder, with properties sensitive to the thermal or magnetic annealing temperatures. 
		We calculate  lattice dynamical properties of Au$_{1-x}$Fe$_x$ alloys 
		using density functional theory methods, where, being a multisite property, reliable interatomic force constant (IFC) calculations in disordered alloys remain a challenge. 
		We follow a two fold approach: (1) an accurate IFC calculation in an environment with nominally zero chemical pair correlations to mimic the homogeneously disordered alloy; and (2) a configurational averaging for the desired phonon properties ({\it e.g.}, dispersion, density of states, and entropy).  We find an anomalous change in the IFC's and phonon dispersion (split bands) near $x$=0.19, which is attributed to the local stiffening of the Au-Au bonds when Au is in the vicinity of Fe.  Other results based on mechanical and thermo-physical properties reflect a similar anomaly: Phonon entropy, e.g., becomes negative below $x$=0.19, suggesting a tendency for chemical unmixing, reflecting the onset of miscibility gap in the phase diagram. Our results match fairly well with reported data, wherever available.
	\end{abstract}
	\maketitle
	{\par} Gold (Au) and iron (Fe) and their alloys continue to attract attention. Due to the higher magnetic state of Fe in Au-Fe than in pure Fe, various properties have been studied, including thickness dependent spin-glass behavior and anomalous hall conductivity in Fe/Au multilayers.\cite{zhang2016anomalous, saoudi2008size, cannella1972magnetic, fritzsche2010loss, kim2003peculiar} Due to its exceptional biocompatibility and favorable physical properties, Au-Fe nanoparticles find various applications in medical sciences, as a
	promising candidate for cancer cell treatment, multimodal magneto-resonance imaging agent, etc..
	\cite{chen2003gold, wu2011cancer, connor2005gold, kukreja2016preparation, amendola2014magneto}

{\par} Gold-rich Au-Fe alloys form a simple face-centered-cubic (fcc) structure. Although fcc is a high-temperature phase, Au-Fe alloys up to 53~at.\%Fe are reported to be easily stabilized at room temperature.\cite{wilhelm2008and, liu2008synthesis, mehendale2010ordered, munoz2013electronic}Due to sensitivity of magnetic and chemical properties to annealing temperatures, these alloys require at most care in their synthesis, especially as disorder is quite common and difficult to control. Hence, chemical disorder plays an important role in their anomalous structural and magnetic properties. With these effects properly understood, thermo-mechanical properties can be suitability tuned for other purposes.

{\par} For alloys to have useful applications, mechanical stability is a necessary criteria. Studying the lattice dynamics provides direct stability information and gives idea about local atomic environment, and related phenomena. Experimentally, techniques like nuclear resonant inelastic X-ray Scattering (NRIXS), inelastic neutron scattering (INS), and Mossbauer spectrometry are used to investigate the elementary excitation in disordered alloys.\cite{munoz2013electronic, lucas2008phonon} But, to date, there is no singularly accepted {\it ab initio}  theoretical  approach available to address lattice dynamics in disordered alloys, mainly due to configurational averaging and the associated computational cost. More precisely, the challenge is to address the off-diagonal disorder arising out of the force constant matrix between two sites. In addition, the diagonal and off-diagonal force constants obey a sum rule that implicitly makes disorder at a site dependent upon its neighborhood, i.e., environmental disorder.

{\par}Historically, various models are proposed to address disorder in some approximation. The Virtual Crystal Approximation (VCA),\cite{nordheim1931electron} and Coherent Potential Approximation (CPA) are two widely known single-site examples, which also suffer from deficiencies.\cite{taylor1967vibrational, renaud1988topological, rowlands2005effects, biava2005systematic, ghosh2002phonons} VCA, the simplest among many, places simple compositional averages of the constituent potentials (clearly physically incorrect for most alloys) and ignores environmental effect, thus neglecting the local distortions. The single-site CPA  captures on-site disorder, but suffers from capturing the multisite effects expected in lattice dynamics, such as, off-diagonal and environmental disorder. Some generalization to the CPA,  {\it e.g.,} Dynamical Cluster Approximation (DCA) \cite{jarrell2001systematic}, and its first-principles version (i.e., non-local CPA (NL-CPA)\cite{rowlands2005effects, biava2005systematic}), or the itinerant CPA (ICPA)\cite{ghosh2002phonons},  address two-site disorder. These methods consist of various promising features, but they are usually limited to specific types of off-diagonal disorder or to small clusters due to exponential computational expense. The Special Quasirandom Structure (SQS) technique\cite{zunger1990special} is being utilized more often to estimate environmental effects of disorder because it supposes a fully-ordered cell (so it can be used within any band-structure method) in a layered arrangement of atoms that nominally exhibits zero chemical pair correlations (within a specified range of neighbor $2-3$ shells) and mimics those of the homogeneously disordered alloy. To predict the lattice dynamical properties of disordered systems, accurate calculation of force constants as well as an appropriate  configurational average over the disorder environment are equally important. 

\begin{table}[h]
	\begin{ruledtabular}
		\begin{centering}
			\begin{tabular}{c c c c c c c c}
				
				x & 1.00 & 0.50 & 0.25  & 0.19  & 0.06 & 0.00 & Direction \tabularnewline
				\hline 
				\vspace{0.05 in}
				Au-Au &  & 26.39 & 21.79 & 19.05 & 16.66 & 17.52 & $110_{xx}$ \tabularnewline
				\vspace{0.05 in}
				Fe-Fe & 9.29 & 14.08 & 9.16 & 2.39 & 9.38 &  & $110_{xx}$ \tabularnewline
				\vspace{0.05 in}  
				Au-Fe &  & 13.89 & 10.98 & 10.98 & 9.20 &  & $110_{xx}$ \tabularnewline
				\vspace{0.05 in}  
				Au-Au &  & 30.57 & 26.18 & 23.06 & 20.33 & 21.03 & $110_{xy}$ \tabularnewline
				\vspace{0.05 in}  
				Fe-Fe & 17.88  & 14.48 & 1.49 & 2.92 & 10.67 &  & $110_{xy}$ \tabularnewline
				\vspace{0.05 in}  
				Au-Fe &  & 16.78 & 12.05 & 13.20 & 10.85 &  & $110_{xy}$ \tabularnewline
				\vspace{0.05 in}  
				Au-Au &  & -6.37 & -6.95 & -6.64 & -6.52 & -5.94 & $110_{zz}$ \tabularnewline
				\vspace{0.05 in}  
				Fe-Fe & 8.15 & 2.68 & -8.01 & -2.35 & -1.60 &  & $110_{zz}$ \tabularnewline
				\vspace{0.05 in}  
				Au-Fe &  & -1.62 & -2.86 & -3.17 & -2.84 &  & $110_{zz}$ 
				
			\end{tabular}
			\par\end{centering}
		\caption{ Force constants (N/m) for Au$_{1-x}$Fe$_x$  along [110]. The measured data for pure Au are 16.63, 20.82, -8.62 along $110_{xx}$, $110_{xy}$, $110_{zz}$ directions respectively.\cite{munoz2013electronic}. }
		\label{final_data}
	\end{ruledtabular}
\end{table}		

	{\par}Here, we combine two techniques to address correctly the above issues: the SQS and Augmented Space Recursion (ASR). ASR is a powerful method to capture multisite disorder effects, as required in the phonon problem. It has been described in great detail in earlier papers.\cite{alam2004vibrational, alam2007phonons} For a given size and symmetry cell, the SQS is used in conjunction with the small displacement method\cite{alfe2009phon}  to calculate the estimated force constants in a disordered alloy. ASR then performs the configurational averaging over a disorder environment, with these disordered force constants. Apart from phonon dispersion, density of states, lifetime, vibrational entropy, we also present a systematic study of thermo-mechanical properties for Au$_{1-x}$Fe$_x$ to explain anomalies.

{\par}Spin-polarized, density functional theory (DFT)\cite{kohn1965self} calculations are performed with a projected augmented wave (PAW)\cite{blochl1994projector} basis within a pseudopotential formalism using the local density approximation (LDA) to the exchange and correlation, as implemented in the Vienna Ab-initio Simulation Package (VASP).\cite{kresse1996efficiency, kresse1999ultrasoft} 
For a different alloy, we have shown the convergence of the phonon dispersion and force constants with SQS cell size, along with k-points. \cite{chouhan2014interplay} 
From this, we chose the optimal 32-atom SQS unit cell\cite{von2010generation} to perform all the calculations for Au$_{1-x}$Fe$_x$ for $x$ = 0.50, 0.25, 0.19 and 0.06, providing good relative accuracy.  For a given SQS arrangement of Au and Fe for each $x$, atoms were relaxed to achieve energy (force) convergence of up to 10$^{-6}$ eV (10$^{-3}$ eV/\r{A}). A high-energy cutoff of 450 eV, with a Monkhorst-Pack $6\times6\times6$ k-mesh grid.\cite{monkhorst1976special}  
For $x$ = 0.50, 0.25, 0.19 and 0.06, respectively, optimized lattice parameters in Au$_{1-x}$Fe$_x$ were 3.83, 3.96, 4.00, and 4.05 \r{A}.
Phonons were calculated using the small displacement method as implemented in PHON,\cite{alfe2009phon} and the  atomic force fields were obtained using 48, 96, 96 and 19 displacements for the respective $x$'s. For elastic constants, we used $PBEsol$ exchange-correlation functional\cite{perdew2008restoring} with a $10\times10\times10$ $\Gamma$-centered k-mesh for total-energy calculation at different strains. 

\begin{figure}
	\includegraphics[width=0.40\textwidth]{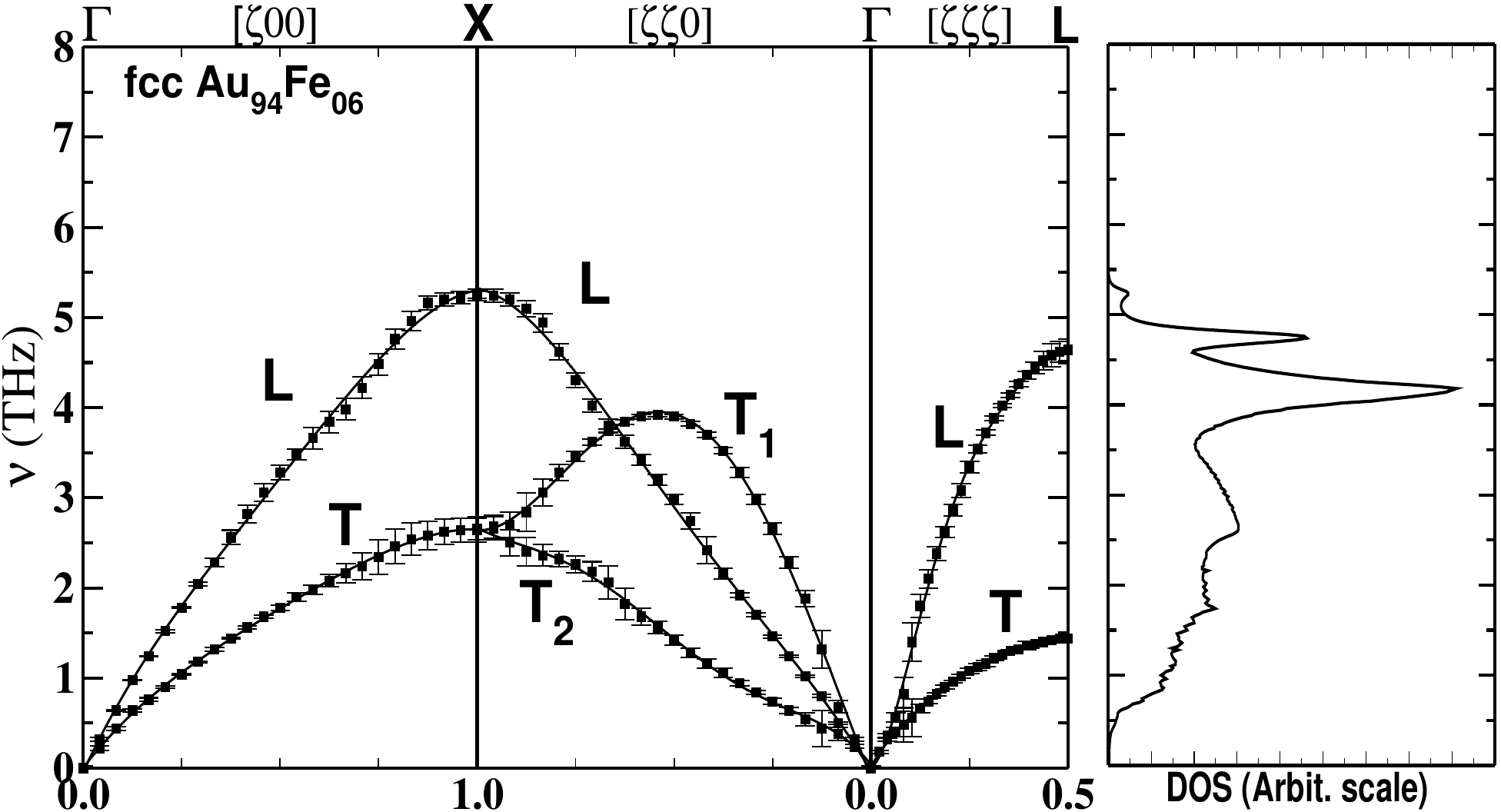}
	\vspace{0.2 in}
	\includegraphics[width=0.40\textwidth]{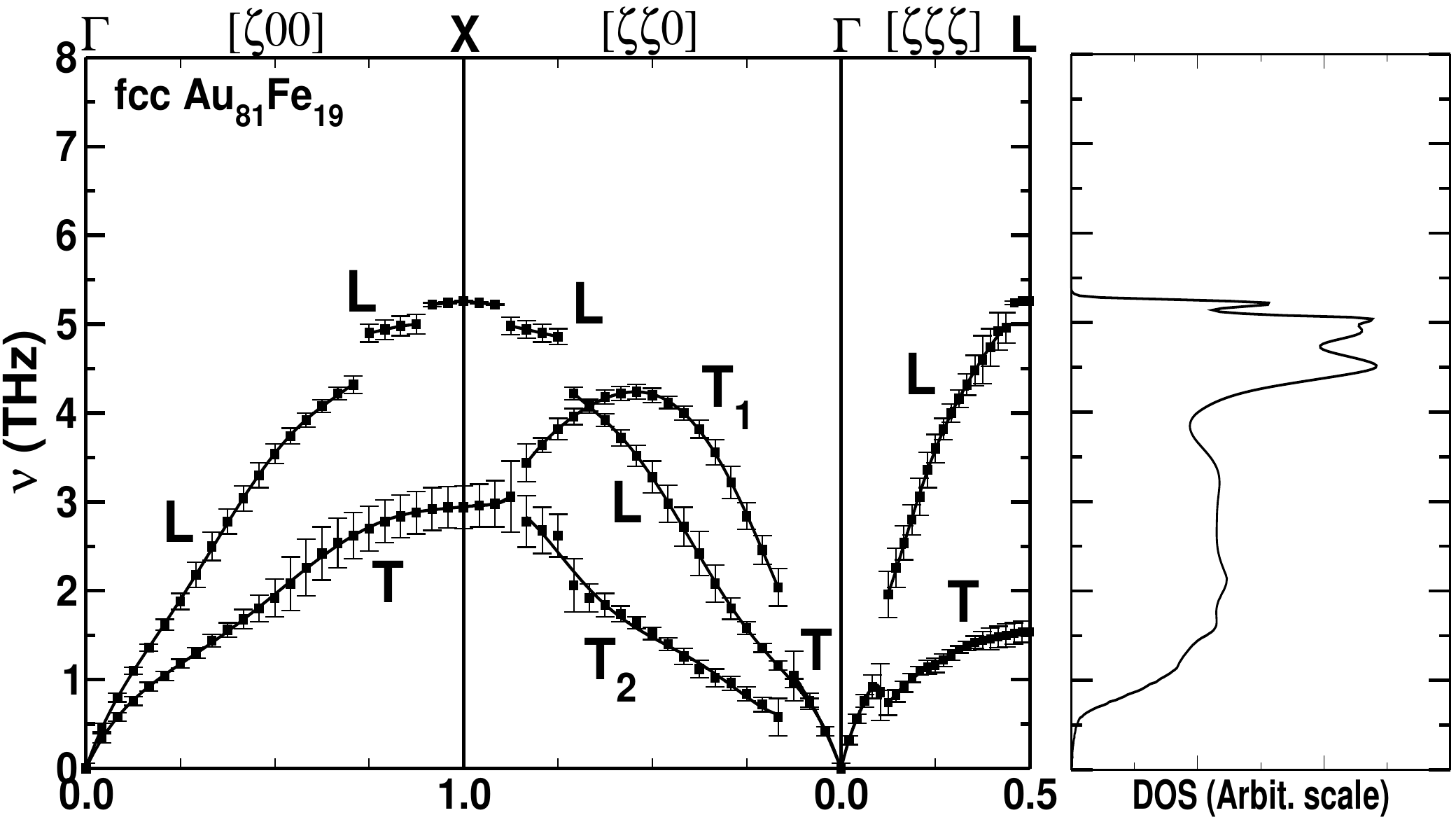}
	\vspace{0.2 in}
	\includegraphics[width=0.40\textwidth]{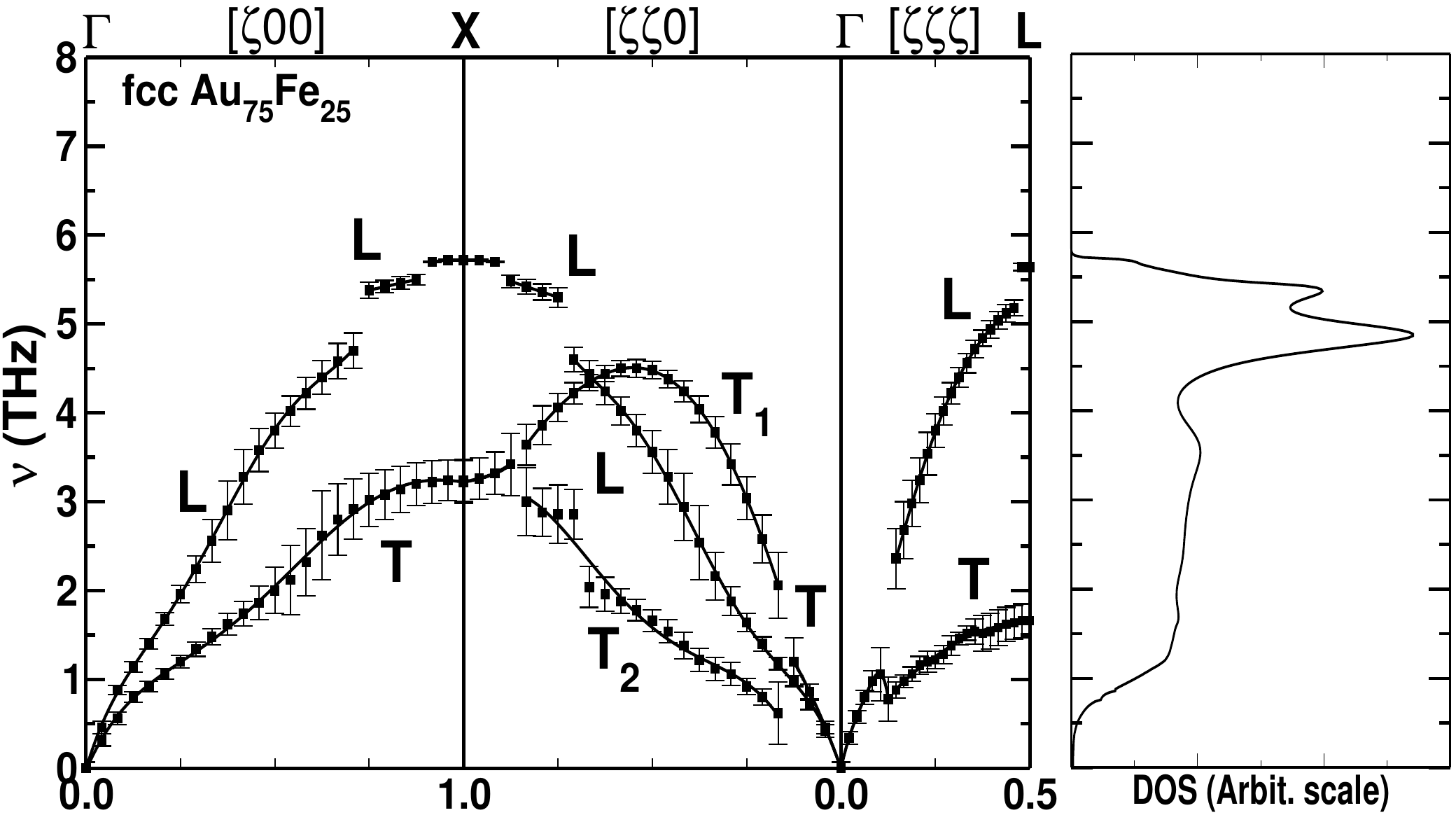}
	\vspace{0.2 in}
	\includegraphics[width=0.40\textwidth]{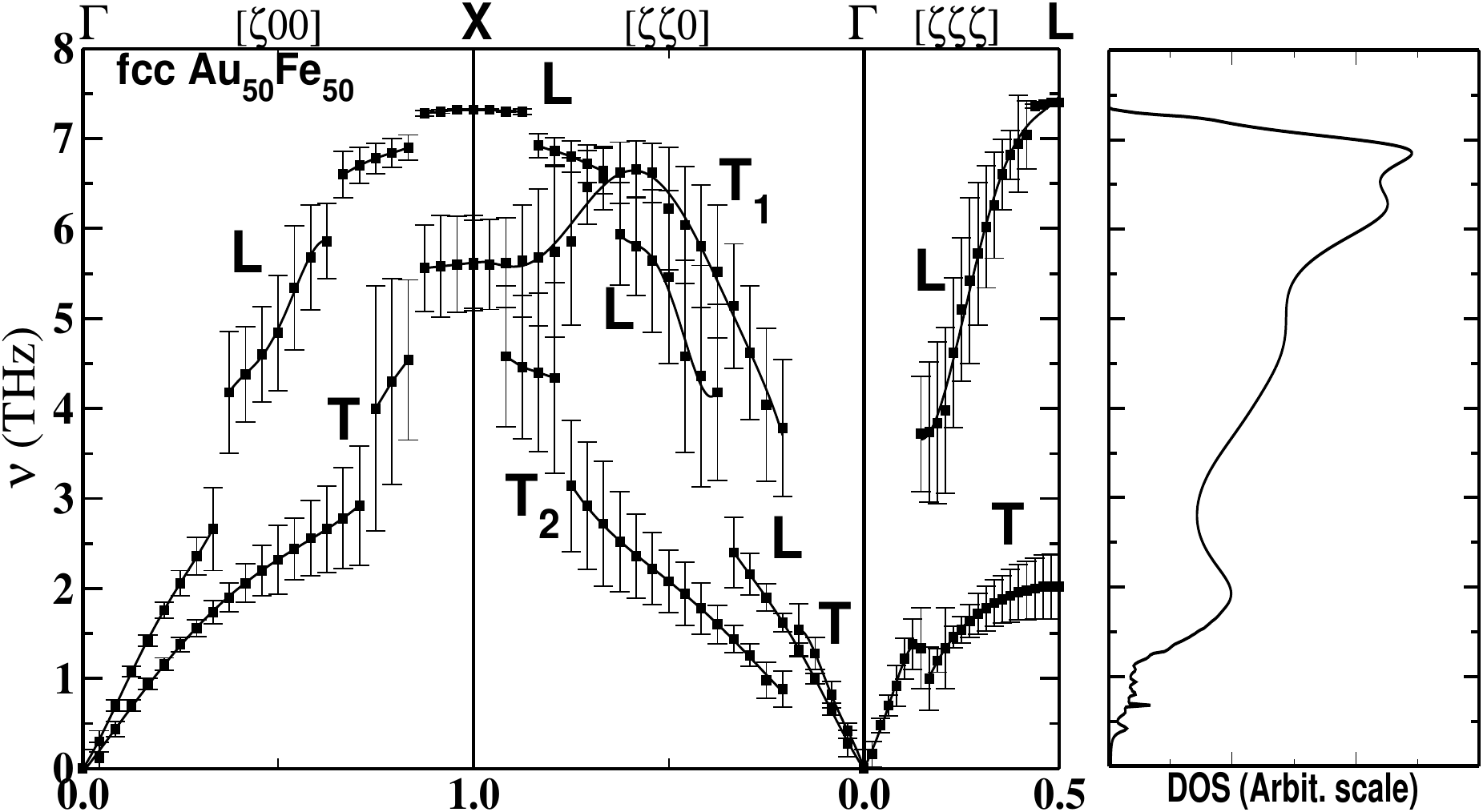}
	\vspace{-0.5cm}
	\caption{(Left) Using force constants from 32-atom SQS, phonon dispersion for Au$_{1-x}$Fe$_x$ along high-symmetry [$\zeta00$], [$\zeta\zeta0$], [$\zeta\zeta\zeta$],  where $\zeta=|\vec{k}|/|\vec{k}_{max}|$ for reciprocal-space vector $\vec{k}$. Longitudinal (L) and transverse (T) modes are indicated. Bars indicate the calculated full width at half maxima (FWHM). (Right) Projected Density of States (DOS).}
	\label{fig1}
\end{figure}

	\begin{table*}[t!]
		\begin{ruledtabular}
			\begin{centering}
				\begin{tabular}{c c c c c c c}
					
					Parameters & Pure Au & Au$_{94}$Fe$_{06}$ & Au$_{81}$Fe$_{19}$ & Au$_{75}$Fe$_{25}$ & Au$_{50}$Fe$_{50}$ & Pure Fe \tabularnewline
					\hline 
					\vspace{0.08 in}
					$C_{11}$ (GPa) & 196.09 $(201.63)^a$ & 192.22 & 222.01 & 205.04 & 211.29 & 328.35 $(243.1)^c$ \tabularnewline
					\vspace{0.08 in}
					$C_{12}$ (GPa) & 164.14 $(169.67)^a$ & 161.38 & 180.94 & 159.24 & 153.41 & 164.56 $(138.1)^c$ \tabularnewline
					\vspace{0.08 in}  
					$C_{44}$ (GPa) & 44.57 $(45.44)^a$ & 48.69 & 57.64 & 61.53 & 81.75 & 136.00 $(121.9)^c$ \tabularnewline
					\vspace{0.08 in}  
					B (GPa) & 174.79 (180.32) & 171.66 & 194.63 & 174.51 & 172.70 & 219.16 (173.1) \tabularnewline
					\vspace{0.08 in}  
					$G_V$ (GPa) & 33.13 (33.65) & 35.38 & 42.80 & 46.08 & 60.61 & 114.36 (94.1) \tabularnewline
					\vspace{0.08 in}  
					$G_R$ (GPa) & 25.97 (26.15) & 26.13 & 33.46 & 36.74 & 47.26 & 107.57 (79.74) \tabularnewline
					\vspace{0.08 in}  
					$G_H$ = $\mu$ (GPa) & 29.55 (29.90) & 30.76 & 38.13 & 41.41 & 53.94 & 110.96 (86.94) \tabularnewline
					\vspace{0.08 in}  
					Y (GPa) & 83.93 (85.01) & 87.07 & 107.37 & 115.12 & 146.56 & 284.82 (223.41) \tabularnewline
					\vspace{0.08 in}  
					C$^{'}$ (GPa) & 15.98 (15.98) & 15.42 & 20.53 $(20.7)^b$ & 22.90 & 28.93 & 81.90 (52.50) \tabularnewline
					\vspace{0.08 in}  
					$C_p$ (GPa) & 119.56 (124.22) & 112.69 & 123.30 & 97.71 & 71.66 & 28.56 (16.18) \tabularnewline
					\vspace{0.08 in}  
					$\nu$ & 0.42 (0.42) & 0.42 & 0.41 & 0.39 & 0.36 & 0.28 (0.28) \tabularnewline
					\vspace{0.08 in}  
					$K_{\zeta}$ & 0.89 (0.89) & 0.89 & 0.87 & 0.84 & 0.81 & 0.63 (0.68) \tabularnewline
					\vspace{0.08 in}  
					$A^Z$ & 2.79 (2.84) & 3.16 & 2.81 & 2.69 & 2.82 & 1.66 (2.32) \tabularnewline
					\vspace{0.08 in} 
					P & 5.92 (6.03) & 5.58 & 5.10 & 4.21 & 3.20 & 1.98 (1.99) \tabularnewline
					\vspace{0.08 in}  
					$\lambda$ (GPa) & 155.09 (160.38) & 151.15 & 169.21 & 146.90 & 136.74 & 145.18 (115.14) \tabularnewline
					\vspace{0.08 in}  
					$v_l$ (m/s) & 3335.20 (3377.67) & 3372.61 & 3735.33 & 3675.10 & 4079.53 & 6523.11 (6058.51) \tabularnewline
					\vspace{0.08 in}  
					$v_t$ (m/s) & 1238.84 (1244.67) & 1282.61 & 1472.16 & 1560.33 & 1915.67 & 3586.33 (3322.86) \tabularnewline
					\vspace{0.08 in}  
					$\Theta_D$ (K) & 162.87 $(162.4\pm2)^d$ & 169.44 & 196.73 & 209.35 & 264.61 & 539.29 $(472.7\pm6)^d$ \tabularnewline
					\vspace{0.08 in}  
					$\kappa_{min}^{Clarke}$ (W/m K) & 0.38 & 0.40 & 0.47 & 0.50 & 0.64 & 1.41  \tabularnewline
					\vspace{0.08 in}  
					$\kappa_{min}^{Cahill}$ (W/m K) & 0.49 & 0.51 & 0.58 & 0.60 & 0.75 & 1.56

				\end{tabular}
				\par\end{centering}
			\caption{Calculated parameters for Au, Fe and four alloys. Parenthetic values are measured data.$^a$\cite{neighbours1958elastic}, $^b$\cite{shiga1986elastic}, $^c$\cite{guo2000gradient}, $^d$\cite{shukla1973calculation}}
			\label{final_data}
		\end{ruledtabular}
	\end{table*}

	{\par}As SQS provides structures with reduced symmetry (not fcc), the force constant matrix become random and asymmetric, which cannot be used directly in ASR for configurational averaging. To extract meaningful parameters for the proper fcc symmetry, a directional average mapping method is adopted. For fcc symmetry, we mapped all 12 nearest neighbor matrix elements for  Au-Au, Fe-Fe and Au-Fe pairs in Au$_{1-x}$Fe$_x$ at each site along [110] using $\phi$$^{101}$ = T$^\dagger\phi^{110}$T, where T is the transformation matrix along different directions.

	{\par} The averaged force constants for all pairs are tabulated in Table I. Notably, Au-Fe force constants become stiffer as we increase the at.\%Fe. Addition of Fe in pure Au makes the Au-Au pair more stiff but Fe-Fe either becomes softer or remains unaffected. Interestingly, Au$_{81}$Fe$_{19}$ shows a turning point, where the force constant matrix elements exhibits a non-monotonous change. This anomaly is also reflected in the phonon dispersion, entropy, and other properties. The origin of this cannot be explained simply by the changes in lattice parameters or the overall electron DOS at the Fermi energy (E$_F$). Below we provide a deeper explanation.

	{\par}Au-Fe alloys are known for their rich variety of magnetic properties.\cite{ling1995origin, wilhelm2008and} Pure Fe in its stable bcc phase has a magnetic moment of 2.13 $\mu_B$/atom, in agreement with previous theoretical and experimental data.\cite{frollani1975magnetic, paduani2006first} As \%Fe decreases from $0.50$ to $0.06$, the Fe moments increases from 2.71 to 2.99 $\mu_B$/atom, also found previously.\cite{munoz2013electronic, sanyal1999electrical} fcc Au is a well known non-magnetic metal. However, we found that the Au 5d moments are  0.083 in Au$_{75}$Fe$_{25}$ and 0.146 in Au$_{50}$Fe$_{50}$, similar to reported theory and experimental values of 0.099 and 0.197, respectively.\cite{wilhelm2008and}  
	
	{\par} Figure \ref{fig1} shows phonon dispersions for  Au$_{1-x}$Fe$_x$ along high-symmetry directions. Notice the split band behavior with  $x\ge19\%$. Such splittings normally arise for systems with dominant mass or force constant disorder. Ni-Pt is a classic example of such behavior. 
	In their elemental phase, Pt-Pt force constants are ~55\% larger than Ni-Ni. Although the force constant difference here is not that significant, the mass difference is higher ($M_{Au}/M_{Fe}=3.53$). Such splitting is a consequence of strong resonance. Near resonances the FWHM become very large, as is clear in Fig.~\ref{fig1}. ASR is expected to correctly address both mass and force constant disorders, mainly in the higher frequency region, as demonstrated in our earlier papers.\cite{alam2004vibrational, chouhan2014interplay} Figure \ref{fig1} also shows the phonon DOS, where the higher (lower) frequency region is dominated by Fe (Au), as expected by mass. It also explains the increase in number of states in the higher frequency region as the \%Fe increases. Our calculated phonon-dispersion and DOS compares fairly well with previous experimental data.\cite{munoz2013electronic} The anomalous band splitting arises for $x\ge$0.19, the turning point in the force constants (Table I). This behavior can be understood from the evolving nature of Au-Au bond in an Fe-matrix. When Fe is substituted in Au, there are two types of force constants that Au-Au pairs acquire. The pairs that do not contain Fe in their vicinity has  force-constants similar to that of pure Au. However, pairs that exists in the neighborhood of Fe increasingly stiffen as \%Fe increase, which causes an increase in the energy of some Au-modes above the cut-off energy of Au-modes and hence causes the splitting. This behavior can be explicitly found in thermo-mechanical properties of the alloy, as seen below. 

	{\par}For a material, thermo-mechanical parameters are directly related to the second-order elastic constants. For a cubic crystal, there are three independent elastic constants denoted by $C_{11}$, $C_{12}$, and $C_{44}$.  Here we use a strain-energy approach\cite{von2010generation, tian2015elastic} to evaluate $C_{ij}$ at various \%Fe, along with bulk modulus (B), shear modulus (G$_V$, G$_R$, G$_H$), Young's modulus (Y), shear constant (C${'}$), Cauchy pressure (C$_P$), Poisson's ratio ($\nu$), Kleinman parameter (K$_\zeta$), Zener's anisotropy ratio (A$^Z$), Pugh's indicator (P), Lames's co-efficients ($\lambda$ and $\mu$), longitudinal and transverse sound wave velocity (v$_l$ and v$_t$), Debye temperature ($\Theta_D$), high-temperature limit to the thermal conductivity, as obtained via Clarke's model ($\kappa_{min}^{Clarke}$)\cite{clarke2003materials} and Cahill's model ($\kappa_{min}^{Cahill}$) \cite{cahill1992lower}. All the properties help asses the mechanical stability of the material (see supplementary material\cite{supplement} for details).
	
	{\par}Table II provides the calculated values of all these properties, including for pure Au and Fe, along with experimental data wherever available.
	Our data (C$_{ij}$'s) agrees within 2-5\% of experimental available data for Au\cite{neighbours1958elastic}. For pure Fe, calculated data differ by more than 25\% with experiment, which is due, as is well-known, to the GGA exchange-correlation function used here for magnetic transition metals. The Born-Huang's mechanical stability criteria\cite{born1955dynamical} ($C_{11}-C_{12}>0, C_{11}+2C_{12}>0$ and $C_{44}>0$) is satisfied for both the elements as well as alloys. 
	The only measured C${'}$ value for Au$_{81}$Fe$_{19}$ is 20.7 GPa, \cite{shiga1986elastic} which compares well with our calculated 20.5 GPa.  The calculated results are expected to be within 10-15\% of the measured values, and will be interesting to be verified experimentally.
	
	{\par}Pure Au has high B but small $G_H$, which makes it very ductile, as seen by its Pugh's indicator (5.92),\cite{pugh1954xcii} where materials with P $>$1.75 are ductile. With increasing \%Fe, the size and coupling force mismatch makes the system more stiffer (see Table I) resulting in reduced ductility and higher Young's modulus.   
	$C_p>0$ \cite{pettifor1992theoretical} suggests metallic bonding character, as well as higher conductivity.  High  $\nu$ values confirm this. $A^Z>1$\cite{zener1947contributions} for both the elements and alloys, points to highly anisotropic deformation in the material (higher possibility of micro-cracks). 
	$K_\zeta$  \cite{kleinman1962deformation} (between 0 and 1) indicates the nature of bonding.  A lower (higher) $K_\zeta$ indicates dominant bond bending (stretching), as found here for Au-Fe. Lame's constant $\lambda$ suggest large incompressibility of these alloys.

{\par}Debye temperature ($\Theta_D$) increases with disorder, attributed to higher mass fluctuation and increase in frequency of thermal vibrational modes. Clarke's model\cite{clarke2003materials} for $\kappa_{min}$ is applicable at a limiting temperature where the dominant phonon wavelength is equal to the interatomic spacing. Whereas, Cahill's model \cite{cahill1992lower} gives the disordered limit of the thermal conductivity. Although total thermal conductivity of Au is higher than Fe, the limiting lattice thermal conductivities show opposite trend (Table II), which remains to be verified experimentally.

{\par}At $x$=0.19, a similar anomaly (as in force constants and phonon dispersion) is encountered in some elastic properties (e.g.,  C$_{11}$, C$_{22}$, B, and C$_P$), reflecting the dominant force constant disorder and the emergence of resonance mode in the dispersion. Such unusual behavior is also predicted by \citeauthor{munoz2013electronic} at $x$=0.2, which are attributed  to the increasing stiffness of Au-Au bonds with increasing \%Fe, and primarily a local effect.

{\par}We have calculated the temperature dependence of excess vibrational entropy, $\Delta S_{vib}=(1-x)\Delta S_{vib}^{Au}+x\Delta S_{vib}^{Fe}$.

\begin{figure}[t]
	\includegraphics[width=0.48\textwidth]{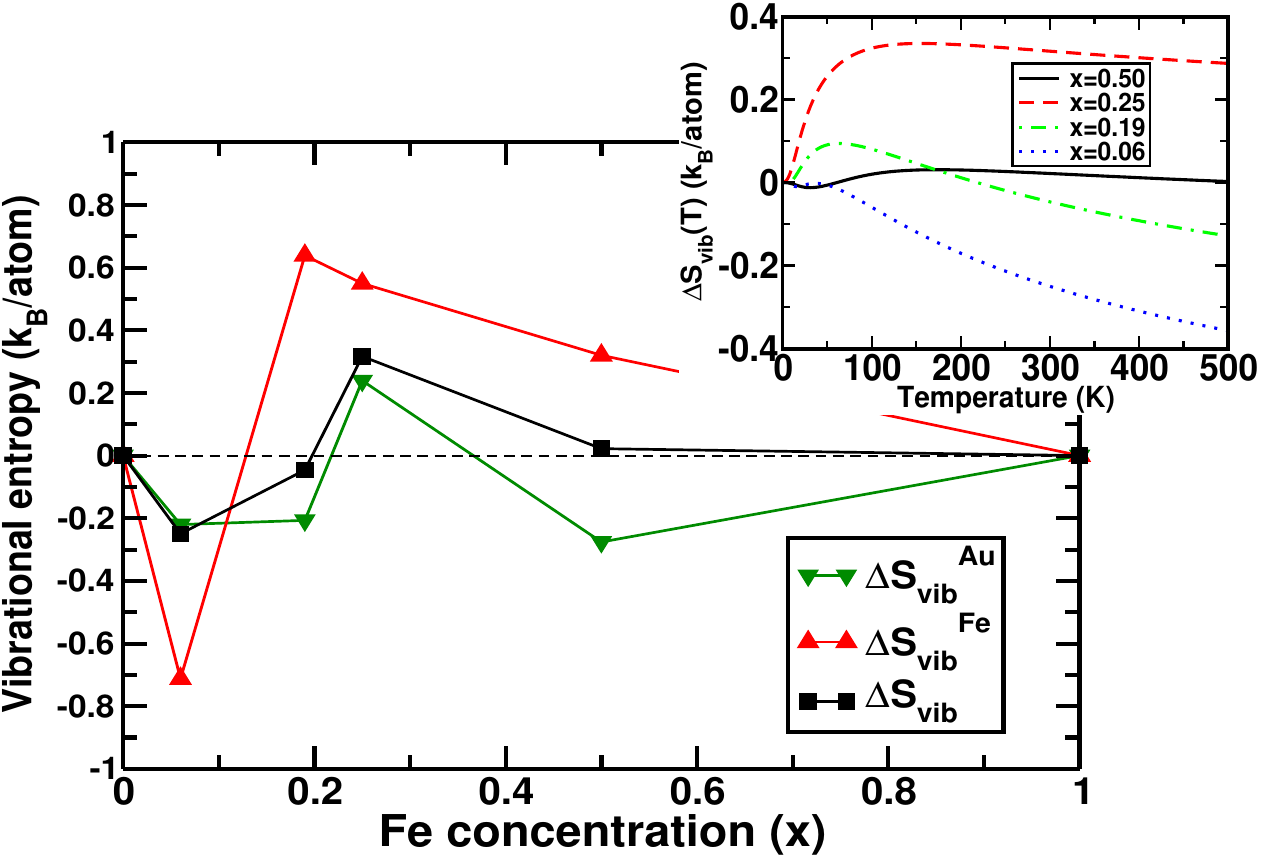}
	\caption{(Color online) Excess phonon entropy at T= 300 K vs. $x$ for Au$_{1-x}$Fe$_x$, and (inset) $\Delta S_{vib}$ vs. T. }
	\label{fig2}
\end{figure}

Here $\Delta S_{vib}^{Au}$ ($\Delta S_{vib}^{Fe}$)  are the partial contribution to vibrational entropy from Au (Fe) respectively at each x.
These are calculated as,
$\Delta S_{vib}^{M}=S_{vib}^{M}(alloy)- S_{vib}^{M}(pure), [M=Au, Fe]$.
$S_{vib}^{M}(alloy)$ is calculated using the partial phonon density of states for each element at a given x. $S_{vib}^{M}(pure)$ is vibrational entropy of pure element M [Au or Fe] in its respective equilibrium phase. Figure \ref{fig2} shows the concentration dependence of excess phonon entropy for Au$_{1-x}$Fe$_x$ at T= 300 K. Square and triangle up (down)
symbol indicate the total entropy and partial entropies for Fe (Au). The inset shows the temperature dependence of total entropy at various Fe concentrations. Clearly $x$=0.19 is an anomalous point which separates the two unique region of phase diagram. In other words, phonon entropy of mixing is negative for Fe concentrations $\le$19\%, beyond which it becomes positive. If we compare the configurational entropy of mixing, the calculated phonon entropy at x=0.06 Fe is much larger and negative in sign. This implies that, upto 19\% Fe, configurational entropy supports chemical mixing, but phonon entropy favours unmixing, predicting miscibility gap in the alloy phase diagram.\cite{massalski1986binary} Such discontinuity in the excess phonon entropy is attributed to the sudden uprise of $\Delta S_{vib}^{Au}$ at x= 0.25, which arises due to the stiffening of Au-Au bonds in the vicinity of the Fe atoms. One can also explain this behaviour from the enhancement of disorder broadening, a known fact to increase the entropy of mixing (as obvious from enhanced FWHM's at x= 0.25). A similar abrupt change in excess entropy is also seen in the temperature dependence of $\Delta S_{vib}$ at $x$= 0.19 (see inset). Our calculated phonon entropy agrees fairly well with similar measured data published elsewhere \cite{munoz2013electronic}. Small discrepancies can be attributed to under estimation of phonon DOS in experimental neutron weighted measurements.

{\par}In conclusion, we employ a new first-principles approach combining the Special Quasirandom Structures (SQS) and Augmented Space Recursion (ASR) formalism to study the lattice dynamical and thermo-physical properties of fcc Au$_{1-x}$Fe$_x$ alloys. This system is interesting because of the large difference in their constituent masses, force constants and scattering lengths. In addition Fe, unlike in its elemental state, acquire larger magnetic moment in the alloy, as such a spin-polarized calculation is performed to accurately predict the IFC's. The phonon dispersion and related data matches fairly well with those reported earlier.\cite{munoz2013electronic}. As the Fe concentration increases, the force constants tends to stiffen in the disordered environment. Above x= 0.19, phonon dispersion shows a split band behaviour suggesting strong resonance often arise due to dominant mass and/or force constant disorder. The anomaly at x= 0.19 is better described from our calculated phonon entropy which suggests the possibility of chemical unmixing below 19\% Fe and hence the onset of miscibility gap in the phase diagram. Such anomaly is also reflected in some of our calculated mechanical properties as well. As Fe concentration increases, size enhancement and coupling force mismatch stiffens the material which accounts for the increased Youngs' modulus and lower ductility. From materials perspective, Au$_{1-x}$Fe$_x$ alloy is predicted to be mechanically stable, very ductile but highly anisotropic (possibility of micro-cracks are high). One of the main ideas of this paper is to establish the combined SQS + ASR approach as an efficient and accurate method to study the lattice dynamical properties for random disordered alloys.

{\par}AA acknowledges DST-SERB (SB/FTP/PS-153/2013) for funding to support this research. JK acknowledges financial support from IIT Bombay in the form of TA-ship.
DDJ was funded by the U.S. DOE, Office of Science, Basic Energy Sciences, Materials Science and Engineering Division. Ames Laboratory is operated for the U.S. DOE by Iowa State University under contract DE-AC02-07CH11358

\end{document}